\documentclass[5p,times]{elsarticle}

\usepackage{booktabs} 
\usepackage{amsmath}
\usepackage{amssymb}
\usepackage{graphicx}
\usepackage{epsfig, boxedminipage}
\usepackage{url} 
\usepackage{algorithmicx} 
\usepackage[ruled]{algorithm}
\usepackage{algpseudocode}
\DeclareGraphicsExtensions{.pdf,.png,.jpg}
\usepackage{multirow}

\journal{Parallel Computing}

\makeatletter
\def\ps@pprintTitle{%
 \let\@oddhead\@empty
 \let\@evenhead\@empty
 \def\@oddfoot{}%
 \let\@evenfoot\@oddfoot}
\makeatother

\begin{document}
\title{MPI+OpenMP Tasking Scalability for Multi-Morphology Simulations of the Human Brain}
\tnotetext[]{DOI: 10.1016/j.parco.2019.03.006. \textcopyright 2019 Elsevier. This manuscript version is made available under the CC-BY-NC-ND 4.0 license \url{http://creativecommons.org/licenses/by-nc-nd/4.0/}}

\author[BSC]{Pedro Valero-Lara}
\author[BSC]{Ra\"ul Sirvent}
\author[BSC]{Antonio J. Pe\~na}
\author[UPC]{Jes\'us Labarta}

\address[BSC]{Barcelona Supercomputing Center (BSC), \{pedro.valero, raul.sirvent, antonio.pena\}@bsc.es}
\address[UPC]{Universitat Polit\`ecnica de Catalunya (UPC), labarta@ac.upc.edu}








\begin{abstract}
The simulation of the behavior of the human brain is one of the most ambitious challenges today with a non-end of important applications. 
We can find many different initiatives in the USA, Europe and Japan which attempt to achieve such a challenging target. In this work, we focus on the most important European 
initiative (the Human Brain Project) and on one of the models developed in this project. 
This tool simulates the spikes triggered in a neural network by computing the voltage capacitance 
on the neurons' morphology, being one of the most precise simulators today. 
In the present work, we have evaluated the use of MPI+OpenMP tasking on top of this framework. 
We prove that this approach is able to achieve a good scaling even when computing a relatively low workload (number of neurons) per node.   
One of our targets consists of achieving not only a highly scalable implementation, but also to develop a tool with a
high degree of abstraction without losing control and performance by using \emph{MPI+OpenMP} tasking. 
The main motivation of this work is the evaluation of this cutting-edge simulation on multi-morphology neural networks. 
The simulation of a high number of neurons, which are completely different among them, is an important challenge.
In fact, in the multi-morphology simulations, we find an important unbalancing between the nodes, mainly due to the differences in the neurons, 
which causes an important under-utilization of the available resources.
In this work, the authors present and evaluate mechanisms to deal with this and reduce the time of this kind of simulations considerably.
\end{abstract}

%
%




\maketitle

\section{Motivation}
\label{motivation}

Today, we can find multiple initiatives that attempt to simulate the behavior of the human brain by computer simulations~\cite{BRAIN, Blue, HBP, OKANO2016}. This is one of the most important challenges in the recent history of computing with a large number of practical applications. The main constraint is being able to simulate efficiently a vast number of neurons (there are about 11 billions of neurons in the human brain) using the current computer technology.    
One of these is the called NEST Initiatives~\cite{NEST}. 
The main motivation of this new initiative is to design and develop a modular brain simulator, which is able to adapt the simulator to the target platform.
In the present paper, the authors focus on \emph{MPI+OpenMP} tasking on homogeneous multi-core clusters. Although we would like to see the work performed in this paper as a new back-end based on \emph{MPI+OpenMP} tasking of the simulator, it is important to note that we are not the developers of such simulator, and the \emph{MPI+OpenMP} tasking is not integrated into such framework.

The present paper extends the previous work~\cite{LaraEuroMPI2018} with additional contributions. Unlike the reference paper~\cite{LaraEuroMPI2018}, this work not only focuses on mono-morphology simulations, where all the neurons share the same shape and size, but it also addresses the problems what arise from the multi-morphology simulations. 
This kind of simulations are more realistic, being one important step forward towards this big challenge. 

One of the most efficient ways in which the scientific community attempts to simulate the behavior of the human brain consists of computing the next 3 major steps~\cite{Diaz16}:
1) the Voltage on neuron morphology, 2) the synaptic elements in each of the neurons and 3) the connectivity between the neurons.
Previously to compute these steps, the network of neurons, the size and shape of the neurons and the connectivity between them are created. 

In this work, we focus on evaluating the \emph{MPI+OpenMP} tasking scalability for the simulation of the human brain. 
Our tests include all the steps of the simulator. 
We describe in detail the numerical model and evaluate the use of \emph{MPI}+\emph{OpenMP} tasking to minimize the impact of the \emph{MPI} communication, and exploit the efficiency of \emph{OpenMP} tasking, not only to orchestrate the overlapping of communication and computation, but also to develop a tool with a high degree of abstraction without losing control and performance.
Additionally, we present and analyze some strategies and implementations on \emph{OpenMP} tasking to deal with the multiple constraints that the multi-morphology simulations present.

One of the most important challenges achieved in the previous work~\cite{LaraEuroMPI2018} was the implementation of an approach that can benefit from the strategies presented in the numerical model (simulator), attaining both, high programming productivity and performance. This was implemented by using the \emph{MPI} function \emph{MPI\_AllGather} for the inter-node communication and \emph{OpenMP} tasking for the intra-node computation and for the overlapping of \emph{MPI} communication and \emph{OpenMP} computation. In the present work, we extend these optimizations to deal with multi-morphology simulations.
Unlike the previous work~\cite{LaraEuroMPI2018}, where all the neurons to be simulated shared the same size and shape, in this work, the authors focus on a much more realistic problem, that is, the simulation of a large number of neurons where all of them are completely different between them, which presents important constraints and challenges in terms of work-distribution.
Both, \emph{MPI\_AllGather} and \emph{OpenMP} tasking can cause an important infra-utilization of the computational resources, in particular on multi-morphology simulations, regarding distributed memory communication and thread scheduling on shared-memory, if the workload is not well distributed among the nodes. 
The authors study and present several optimizations, using \emph{OpenMP} tasking and different ways to sort the neurons and the assignation of these to the computational
nodes, to deal with this important problem without losing the achievements presented in the previous work; keeping a high programming productivity and scalability. 

This paper is structured as follows. Section~\ref{simulator} describes the physical problem at hand and the general numerical framework that has been
selected to cope with it.
In Section~\ref{parallel}, we present the specific parallel features for the resolution of mono-morphology simulations, 
as well as the parallel strategies envisaged to optimally enhance the performance. We also include the performance study 
to evaluate the scalability of this approach. 
Section~\ref{parallel-multi} extends the study presented in the previous section on multi-morphology simulations. 
Section~\ref{related} presents the state-of-the-art references and related work.
Finally the conclusions are outlined in Section~\ref{conclusions}.

\section{Human Brain Simulator}
\label{simulator}

The simulator is divided into two major tasks: i) computation on neurons' morphology (voltage capacitance and spikes triggered), and ii) exchange of the spike events between neurons which are connected through synapses. 
In the model used by the simulator, the neurons can be seen as multi-compartment cables~\cite{NESTMC-paper} composed of active electrical
elements. This model can benefit from several HPC capabilities such as vectorization, tasking, and overlapping of \emph{MPI} communication and \emph{OpenMP} computation. 

\begin{figure*}[h]
\centering
\includegraphics[width = 0.95\textwidth]{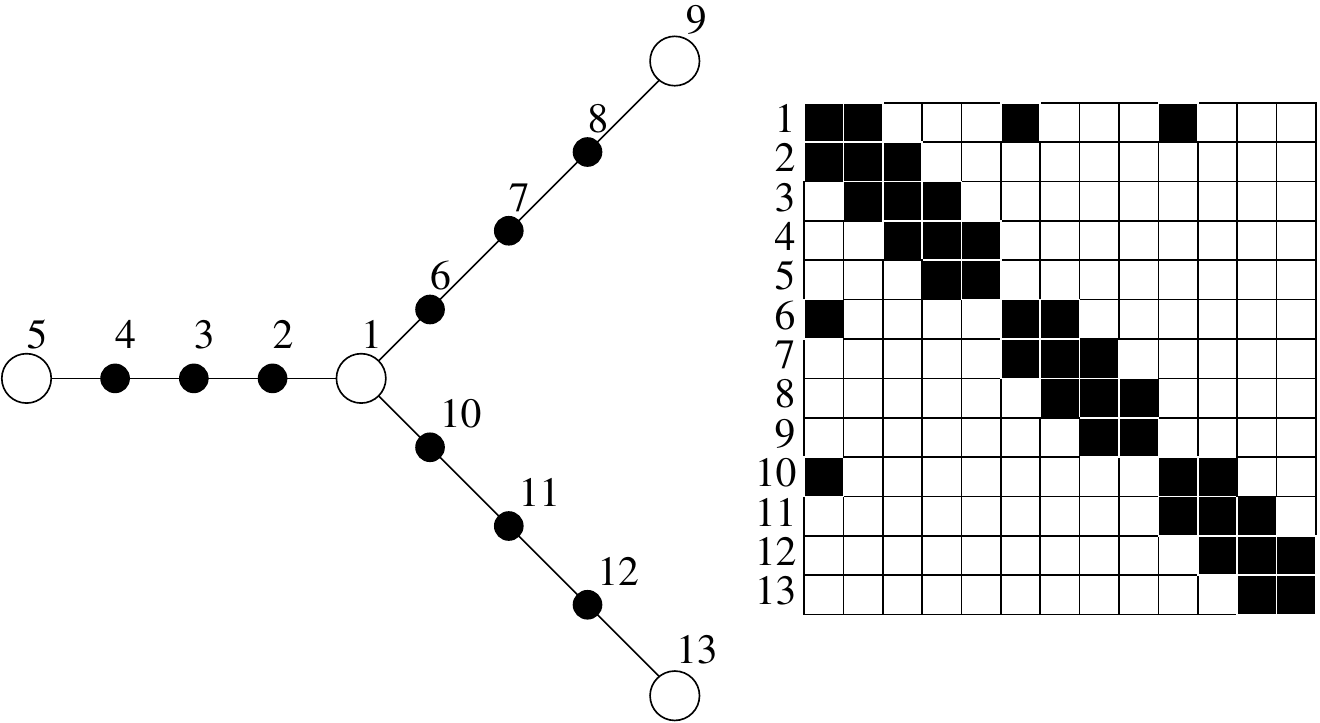}
\caption{Example of a neuron morphology (left) and its sparsity pattern mapped on a matrix (right).}
\label{s:fig2}
\end{figure*}

Next we describe the numerical framework behind the computation of the voltage capacitance on neurons morphology~\cite{Ben2010}, which is one of the most time consuming steps of the simulation.
It follows the next general form:

\begin{equation}
C\frac{\partial V}{\partial t} + I = f \frac{\partial}{\partial x}(g\frac{\partial V}{\partial x})
\end{equation}

where \emph{f} and \emph{g} are functions on x-dimension and the current \emph{I} and capacitance \emph{C}~\cite{Diaz16} depend on the 
voltage \emph{V}.
Discretizing the previous equation on a given morphology (as the simple morphology shown in Figure~\ref{s:fig2}) we obtain a system that has to be solved every time-step.
This system must be solved at each point:

\begin{equation}
a_iV_{i+1} + d_iV_i + b_iV_{i-1} = r_i
\end{equation}

where the coefficients of the matrix are defined as follows:

\begin{center}
upper diagonal: $a_i=-\frac{f_ig_{i+\frac{1}{2}}}{2\Delta_x^2}$ \\
lower diagonal: $b_i=-\frac{f_ig_{i+\frac{1}{2}}}{2\Delta_x^2}$ \\
diagonal: $d_i=\frac{C_i}{\Delta_t}-(a_i+b_i)$ \\
rhs: $r_i=\frac{C_i}{\Delta_t}V_i-I-a_i(V_{i-1}-V_i)-b_i(V_{i+1}-V_i)$
\end{center}

The $a_i$ and $b_i$ are constant in time, and they are computed once at start up.
Otherwise, the diagonal ($d_i$) and right-side-hand (rhs) coefficients are updated every time-step when
solving the system.

The discretization explained above is extended to include \emph{branching}, where the spatial
domain (neuron morphology) is composed of a series of one-dimension \emph{branches} that are joined at node points 
according to the neuron morphology.

For the sake of clarity, we illustrate a simple example of a neuron morphology in Figure~\ref{s:fig2}, composed of 3 branches and one node, which connects the branches.
It is important to note that the graph formed by the neuron morphology is an acyclic graph, i.e. it has no loops.
The nodes are numbered using a scheme that gives the matrix sparsity structure that allows to solve the system 
in linear time. 
 
To describe the sparsity of the matrix from the numbering used, we need an array $p$ ($p_i~i\in[2:n]$) which stores the parent indexes of each node.
The pattern of the matrix which illustrates the morphology is also graphically illustrated in Figure~\ref{s:fig2}. This kind of matrices are known as Hines matrices.


The Hines matrices~\cite{Hines84,Conte80} feature the following properties: they are symmetric, the diagonal coefficients are all nonzero and 
per each off-diagonal element, there is one off-diagonal element in the corresponding row and column.

Given the aforementioned properties, the Hines systems, i.e. $Ax=b$ where $A$ is a Hines matrix, can be efficiently solved by using an algorithm similar to Thomas algorithm 
for solving tridiagonal systems. This algorithm, called Hines algorithm, is almost identical to the Thomas algorithm except by the sparsity pattern given by the morphology 
of the neurons whose pattern is stored by the $p$ vector.
An example of the sequential code used to implement the Hines algorithm is illustrated in pseudo-code in Algorithm~\ref{Halgorithm}.

\begin{algorithm}[h]
\caption{Hines algorithm.}
\label{Halgorithm}
\begin{small}
\begin{algorithmic} [1]
\State void solveHines(double *a, double *b, double *c, 
\State ~~~~~~~~~~~~~~~~~~~double *rhs, int *p, int n) 
\State {\bf// a $\rightarrow$ upper vector, b $\rightarrow$ diagonal vector, c $\rightarrow$ lower vector}
\State {\bf// Backward Sweep}
\For{$i=n-1 \to 0$}
	\State factor~~~~= a[i] / b[i]
        \State b[{\bf p[i]}]~~~-= factor $\times$ c[i]
        \State rhs[{\bf p[i]}] -= factor $\times$ rhs[i]
\EndFor
\State rhs[0] /= b[0];
\State {\bf// Forward Sweep}
\For{$i=1 \to n-1$}
	\State rhs[i] -= c[i] $\times$ rhs[{\bf p[i]}]
	\State rhs[i] /= b[i] 
\EndFor
\end{algorithmic}
\end{small}
\end{algorithm}

Once the voltage is computed, we compute the spikes. Basically, this consists of going through the different points on the neurons' morphology where there is a synapse (a connection between two neurons) and check if the voltage in these points is higher or lower than a given threshold to trigger or not a spike.  

It is important not to forget one of the most important challenges into this model. This is the massive spike exchange between the neurons, which can be a problem on current distributed memory clusters due to the large difference between communication and computation speed, in particular if this communication has to be carried out using the \emph{MPI\_AllGather} routine, since one neuron can be connected (through the synapses) with a huge number of neurons. 
The strategy followed in this model to deal with this problem consists of the next ideas.
The simulation time is divided into two different time-step factors, one local and one global (Figure~\ref{s:fig4}).
In every local (\emph{dt}) iteration, all the neurons are computed and the spike events are stored in one local buffer. 
There is no \emph{MPI} communication at this level. After computing several local steps, we compute what we call a global (\emph{Network delay}) step. The MPI (\emph{MPI\_AllGather}) communication is carried out in this step. Basically, it consists of two tasks: first, we store the spike events triggered, along one global step, into another local buffer; then, we send/receive the information of the spikes to/from the rest of nodes using \emph{MPI\_AllGather}. In this way, all the nodes have the information about the spikes triggered along the simulation.


It is important to note that, although this model is in need of exchanging the information about the spikes at every global step, there is an important difference between the data size sent in the spikes' exchange (\emph{MPI\_AllGather}) and the operations computed along the local steps that compose one global step. For instance, let us assume that we have \emph{N} neurons of size \emph{M}, where every neuron has \emph{S} synapses. Let us also assume that we compute a total of local steps equal to \emph{D} per global step, and along these steps $S_p$ spikes were triggered ($S_p <= S$). The operations performed every global step are $ D \times ( N \times ((8 \times M) + (S)))$, where $( 8 \times M )$ corresponds to the necessary operations to compute the Hines algorithm on a neuron of size \emph{M}, and \emph{S} corresponds to the operations performed for spike computation on synapses. While the data transferred is $N \times S_p$, the operations computed are $D \times N \times (8 \times M)$, which are much more numerous than the data transferred. Depending on the simulation, these parameters can be very different; however, commonly used value~\cite{Valero-LaraICCS17,valero2018} for \emph{M} are about $10 - 800$, for \emph{S} is $200- 1000$, and \emph{N} depends more on the hardware (memory) limit and simulation time desired than on a specific range, but in our experiments we execute in the range of dozens to hundreds of thousands of neurons. \emph{D} can be about of $10-20$.

\subsection{Overlapping Computation \& Communication}

In this section we will explain the characteristics of the model implemented in the simulator, which allows the overlapping of computation and communication. 
The communication is carried out every global step (Figure~\ref{s:fig4}). The data transferred are the spike events triggered along one global step. 
This information is first stored locally, in each of the nodes.
All the nodes must share this information with the rest of nodes to know the necessary statistics of the simulation, such as the source and destination of such spikes.
This information does not influence on the rest of the steps of the simulation, so this step can be performed at the same time that the others are being computed.

To keep updated the information regarding the spikes triggered among all the nodes, two buffers are used per node: one buffer, which stores the spikes triggered by the neurons computed by each of the nodes with the \emph{local} information of the simulation (for the sake of clarity we call this buffer as \emph{local buffer}), and one buffer (\emph{global buffer}), which is updated after every global step with the information that come from the other nodes via \emph{MPI\_AllGather} with the \emph{global} information of the simulation. 

Although this communication can be overlapped with the computation into one global step, it is not possible to perform this overlapping between different global steps.
Due to this, we can see this communication as a ``synchronization point''. Hence, after computing one global step, we cannot start the execution of the next global steps, until the information of the triggered spikes has been shared among the nodes. To avoid important and potential infra-utilization of resources due to this ``synchronization point'', it is important that all the nodes have a similar workload. 
This can be partiality solved by overlapping communication with computation. Although the data transferred among nodes is not large, a load unbalance among nodes together with the communication (synchronization) can hinder the performance considerably.
As shown in the following sections, some of the proposed optimizations have been implemented to achieve load balancing among nodes, in particular for multi-morphology simulations.

\section{Parallel MPI+OpenMP Tasking Simulator on \\ Mono-Morphology Simulations}
\label{parallel}

After reviewing the main characteristics of the simulator, we focus on its parallelization. We decided to use both, \emph{MPI} and \emph{OpenMP}, since both are standards, and the most extended and used programing models for distributed memory and shared memory computation, respectively. In fact, in the last years the concept of \emph{MPI+X} is more and more popular, being the \emph{OpenMP} standard the most popular and widely used candidate for the \emph{X} unknown into the equation. 

As commented above, for \emph{MPI} call, the simulator makes use of the \emph{MPI} instruction, \emph{MPI\_AllGather}. This is because of the particular nature of the target application (see Section~\ref{simulator}). Although this routine is among the least scalable routines in \emph{MPI}, this simulator is able to achieve good scalability by minimizing the cost of this routine thanks to both, the model used for the simulation and the parallelization implemented.

Recently, since \emph{OpenMP 3.0}~\cite{OpenMP}, it is possible to use tasking into \emph{OpenMP}. Using tasking not only allows us to declare the dependences among tasks and let the compiler deal with the best distribution of the tasks on multi-core processors, but it also helps us to implement \emph{MPI+OpenMP} codes very easily. For instance, we can encapsulate \emph{MPI} routines into \emph{OpenMP tasks}, which considerably simplifies the interoperability between both standards and the overlapping of \emph{MPI} communication with \emph{OpenMP} computing. 

\begin{figure*}[t]
\centering
\includegraphics[width = 1\textwidth]{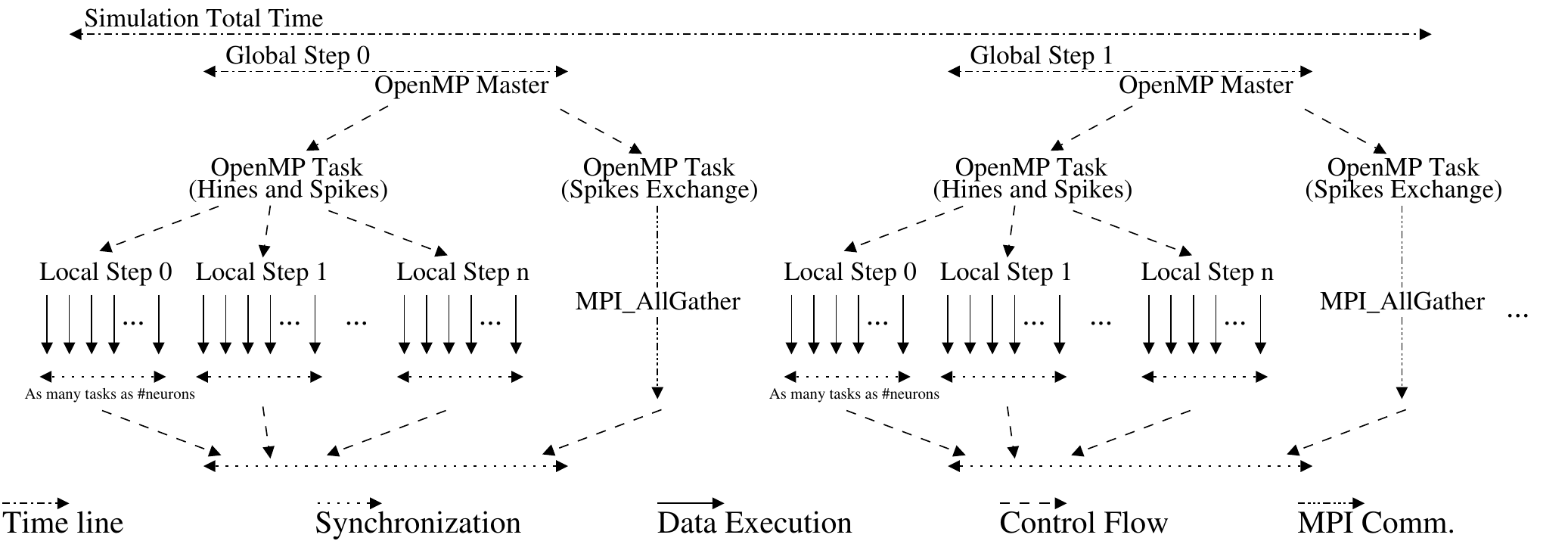}
\caption{Parallel model implemented for the simulator based on MPI+OpenMP tasking.}
\label{s:fig4}
\end{figure*}

Our target is to achieve the model graphically illustrated in Figure~\ref{s:fig4}, overlapping the computation on the neurons' morphology (Hines and Spikes in Figure~\ref{s:fig4}) and the \emph{MPI} communication (Spikes Exchange in Figure~\ref{s:fig4}). Although there are many different ways to achieve this target, our target is to perform an ``easy to implement'' and as much transparent (from the programmer's point of view) as possible approach that can yield a good scalability and performance. It is also important that our implementation makes use of standard programing models, minimizing (even avoiding) the programing effort to maintain, port and/or tune our application. Keeping this idea in mind, we parallelize our code using \emph{OpenMP} tasking in the way that is illustrated by Algorithm~\ref{Pimplementation}.

\begin{algorithm}[h]
\caption{Parallel implementation based on MPI + OpenMP tasking.}
\label{Pimplementation}
\begin{small}
\begin{algorithmic} [1]
\State while(global\_step $\leq$ total\_steps)\{
\State \hspace{0.5cm}{\bf\#pragma omp parallel}\{ 
\State \hspace{0.5cm}{\bf\#pragma omp master}\{
\State \hspace{1cm}{\bf\#pragma omp task}
\State \hspace{1cm}\emph{MPI\_AllGather(Spikes);}
\State \hspace{1cm}{\bf\#pragma omp task}
\State \hspace{1cm}for( local\_step=0; local\_step $\leq$ D; local\_step++)\{
\State \hspace{1.5cm}for( i=0; i $\leq$ \#Neurons; i++)\{
\State \hspace{2cm}{\bf\#pragma omp task}\{
\State \hspace{2cm}\emph{Hines(Neuron[i]);}
\State \hspace{2cm}\emph{Spikes(Neuron[i]);}
\State \hspace{2cm}{\bf\}} //End omp task
\State \hspace{1.5cm}\} //End for \#Neurons
\State \hspace{1.5cm}{\bf \#pragma omp taskwait}
\State \hspace{1cm} \} //End for D 
\State \hspace{0.5cm}{\bf\}} //End omp master
\State \hspace{0.5cm}{\bf\}} //End omp parallel
\State \hspace{0.5cm} global\_step++;
\State \}
\end{algorithmic}
\end{small}
\end{algorithm}
 
The parallelization is based on \emph{OpenMP pragmas}. First we open (fork) a \emph{parallel region} by using \emph{\#pragma omp parallel} every \emph{global} time-step (\emph{global\_step} in Algorithm~\ref{Pimplementation}). After this, since we use \emph{OpenMP} tasking, we must use \emph{\#pragma omp master}. Note that although the use of \emph{\#pragma omp master} could be substituted by the use of \emph{\#pragma omp single \ldots nowait}, the \emph{master} construct is faster and has a better integration with \emph{MPI}. At this level, we can create as many \emph{OpenMP} tasks as we want.
We also take advantage of using \emph{OpenMP} nesting for the modularity of the code, where every major step (\emph{MPI} communication, Hines and Spikes computation) has been implemented in separate files. 
So, while in the fist level (file) we have the \emph{while} loop show in line~1 of Algorithm~\ref{Pimplementation}, the exchange of spikes and the computation of Hines algorithm on neurons' morphology are implemented in separate files.
The use of both, \emph{OpenMP} tasking and nesting, allows us to do this without important modifications in the code.
Also, as we will see in the following sections, the use of \emph{OpenMP} tasks in the computation of Hines and Spikes (lines 10 and 11 in Algorithm~\ref{Pimplementation}) instead of other \emph{OpenMP} constructs, such as \emph{parallel for} or \emph{taskloop}, gives us a better flexibility and control that allows us to identify and explore novel approaches, which have helped to accelerate the simulation.
Both, the number of neurons and the computational cost per neuron (task) are large enough to make an effective use of \emph{OpenMP} tasking. 
However, for mono-morphology simulations, the use of these \emph{OpenMP} constructs can offer a good performance (as the workload in every iteration of the loop is the same) for multi-morphology simulations (where the computational cost of every iteration of the loop is completely different) the use of the \emph{parallel for} or \emph{taskloop} \emph{OpenMP pragma} is too rigid to achieve a good balancing~\cite{Valero-LaraPDP18}. In fact, as shown in the next sections, \emph{OpenMP} tasks are able to balance this kind of non-homogeneous loops in a simple and elegant way.
Furthermore \emph{OpenMP} tasks can help for a better and deeper performance analysis (Section~\ref{performance}), since the analysis can be performed at very low granularity (task-level granularity). 

In the first level of parallelism, we use two tasks, one for \emph{MPI} communication and one for \emph{OpenMP} computation. The last task (line 6 in Algorithm~\ref{Pimplementation}) instantiates one task (line 9 in Algorithm~\ref{Pimplementation}) per neuron, which computes the \emph{Hines} algorithm and the \emph{Spike} computation on one particular neuron. After the \emph{\#Neurons} for loop (line 14 in Algorithm~\ref{Pimplementation}), we must synchronize the tasks instantiated because of the data-dependences among different iterations of the \emph{D} for loop. 
This additional level of parallelism (nesting) using \emph{OpenMP} tasks eases the overlapping between the two main steps, the spikes exchange and the computation of the voltage capacitance.
The parallel region is closed (join) every \emph{global} time step. As we show in the next section, the use of fork-join, nesting, and tasks synchronization does not represent an overhead due to the computational intensity to compute the Hines algorithm on a high number of neurons. 
   
As shown in Algorithm~\ref{Pimplementation}, ``by only'' using 5 different \emph{OpenMP} pragmas and one \emph{MPI} primitive, we are able to have a portable, easy to maintain and optimized code for the simulation of the human brain.

\subsection{Analysis of Scalability}
\label{performance}

The platform used in our experiments is a cluster composed of $39$ NUMA nodes with 2 sockets each,
using Intel Xeon CPU E5649, see Table~\ref{tab.platform1} for more details.
Hyperthreading is not enabled. 

\begin{table}[t]
  \caption{\label{tab.platform1}
    Details of the architecture used}
  \begin{center}
\begin{tabular}{c c}
\textbf{Platform}       & Xeon E5649 (Westmere) at 2.53 GHz \\ \hline \hline
\textbf{Cores}          & 2$\times$6 \\
\textbf{On-chip Memory} & L1 32KB (per core) \\
                        & L2 256KB (per core) \\
             		& L3 12MB (unified) \\
\textbf{Main Memory}    & 24GB DDR4 \\
\textbf{Compiler}       & gcc 6.2.0 \\
                        & openmpi v3.0.0 \\
\textbf{Network}        & 2 Infiniband QDR (4 Gbit/s each) \\ 
			& non-blocking network \\
\end{tabular}
\end{center}
\end{table}

\begin{figure}[t]
\centering
\includegraphics[width = 0.5\textwidth]{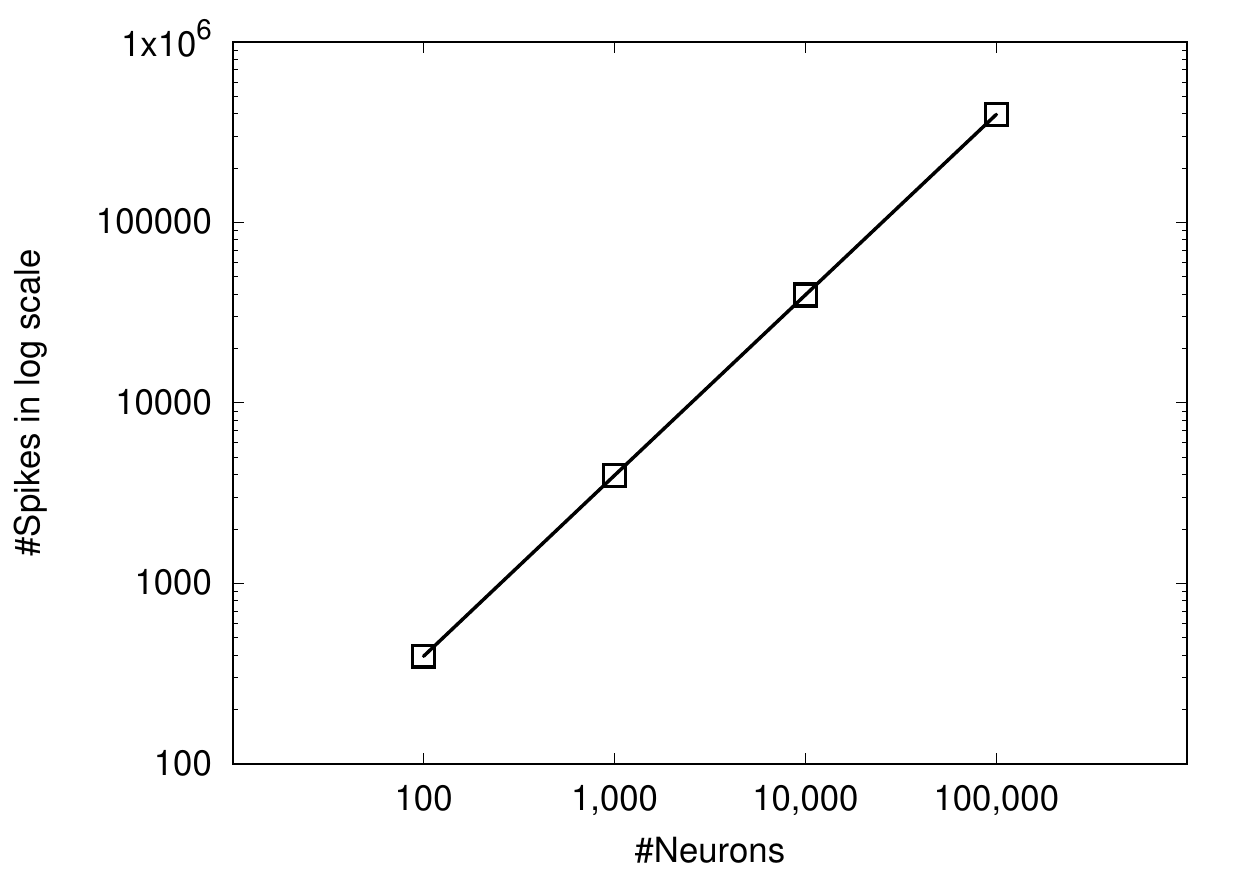}
\caption{Total number of spikes triggered in each of the experiments.}
\label{p:fig-spikes}
\end{figure}

In all the experiments we use one \emph{MPI} process per node, and as many \emph{OpenMP} threads as cores available (12 in our test platform). 
We use the default values for the parameters regarding the size of the neurons (450) and number of synapses per neuron (500)~\footnote{For those experiments, which involve 100 neurons, the number of synapses per neuron is 100, achieving a fully-connected neural network}. The simulations consist of computing 10 global iterations and 10 local iterations per global iteration. The number of spikes triggered along the execution depends only on the parameters of the simulation, not on the number of nodes and cores per node. In particular it has a strong relationship with the number of neurons computed. In general, the more neurons, the more spikes are generated (see Figure~\ref{p:fig-spikes}). 
The number of spikes triggered (data size transferred) can be different at every global iteration. 


\begin{figure}[t]
\centering
\includegraphics[width = 0.5\textwidth]{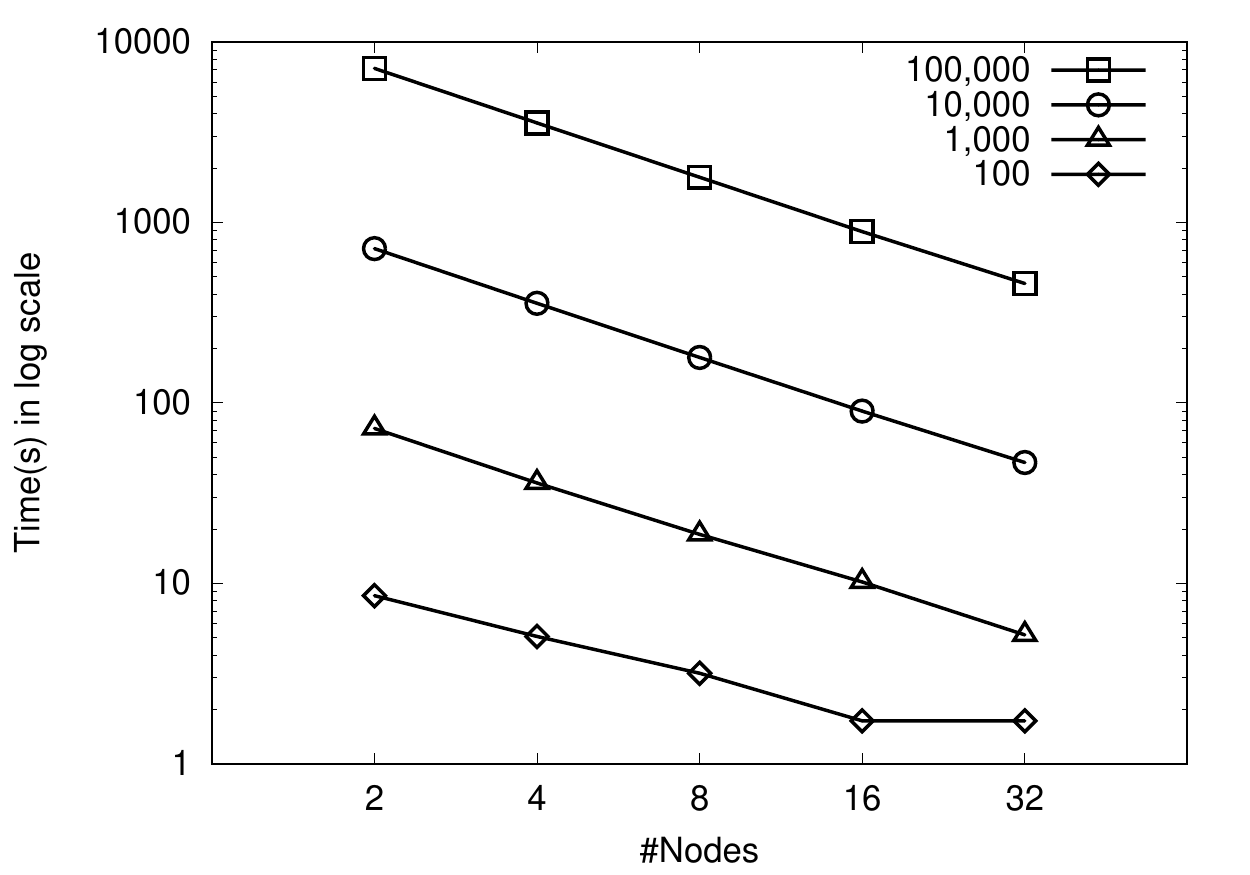}
\caption{Strong scaling analysis depending on number of neurons.}
\label{p:fig2}
\end{figure}

Next, we analyze the strong and weak scaling of our approach. First we focus on strong scaling analysis. In this case, we have launched 4 test cases regarding the number of neurons computed (100, 1,000, 10,000 and 100,000). Figure~\ref{p:fig2} graphically illustrates the strong scaling analysis by increasing the number of nodes keeping constant the number of neurons computed in the simulation. The neurons computed per node depend on the number of nodes used. For instance, for a simulation composed of 100,000 neurons and executed on 2 nodes, half of the neurons (50,000 neurons) are computed on one node and the rest of neurons on the other node. In case of using 4 nodes, every node computes 25,000 neurons.   
As shown, our approach is able to achieve an ideal strong scaling, except in the case of computing 100 neurons on 32 nodes (384 cores), where the \emph{MPI} communication dominates against the computations on the neurons. This is because, in this case, the number of neurons per node is very low (3). Unlike the previous scenario, when computing 1,000 neurons on 32 nodes (31 neurons per node), we do not see this behavior, so that it is proven that it is not necessary to have a high workload per node to yield good scaling. 

\begin{figure}[t]
\centering
\includegraphics[width = 0.5\textwidth]{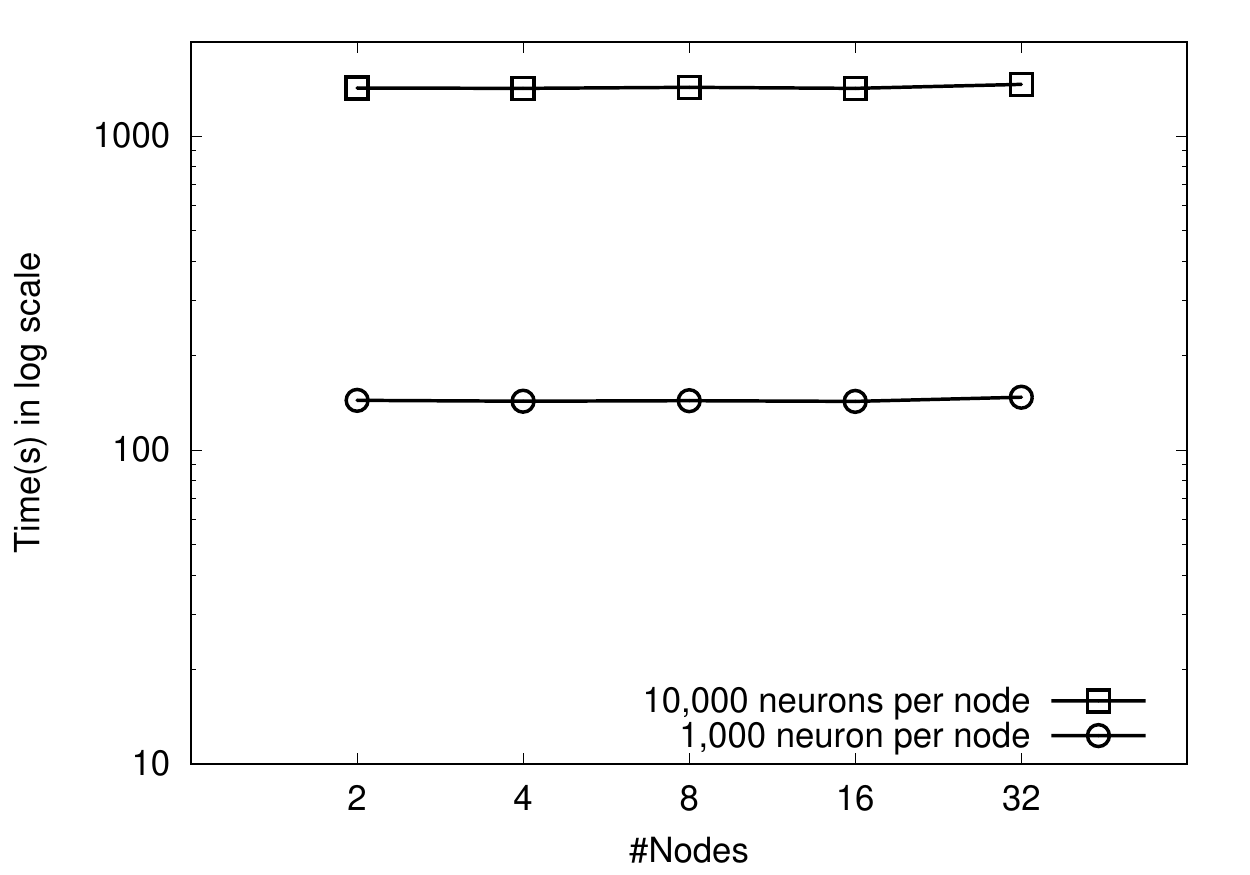}
\caption{Weak scaling analysis.}
\label{p:fig3}
\end{figure}

After analyzing the strong scaling, we analyze the weak scaling by increasing the number of nodes while keeping constant the number of neurons computed in the simulation per node. In this case, we have two test cases regarding the number of neurons distributed per node, 1,000 and 10,000 neurons per node, respectively. As shown in Figure~\ref{p:fig3}, our \emph{MPI+OpenMP} tasking implementation is able to achieve ideal weak scaling.  

\subsection{Detailed Performance Analysis}
\label{performance2}

In order to perform a deeper analysis on the implementation presented, we have used the tools \emph{Extrae+Paraver}~\cite{Llort13}. \emph{Extrae} is a dynamic instrumentation tool to trace programs compiled and run using \emph{OpenMP}, \emph{OmpSs}, \emph{pthreads}, \emph{MPI} or a combination of the previous programming models (different \emph{MPI} processes using \emph{OpenMP} threads within each \emph{MPI process}). \emph{Extrae} generates trace files that can be later visualized with \emph{Paraver}.


\begin{figure*}[t]
\centering
\includegraphics[width = 0.85\textwidth]{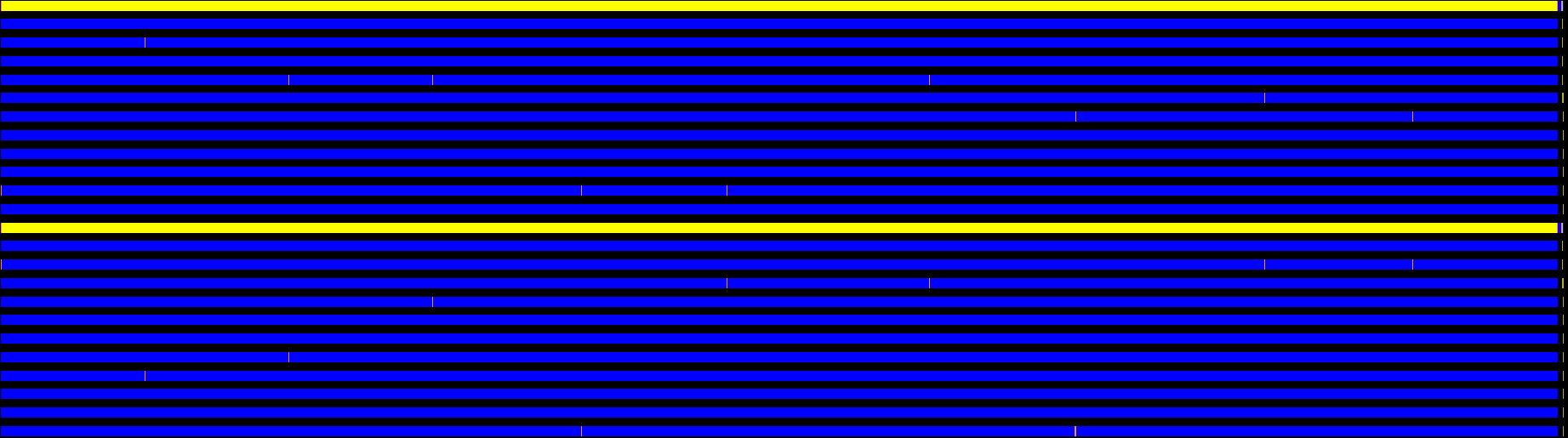} \\
\vspace{0.5cm}
\includegraphics[width = 0.85\textwidth]{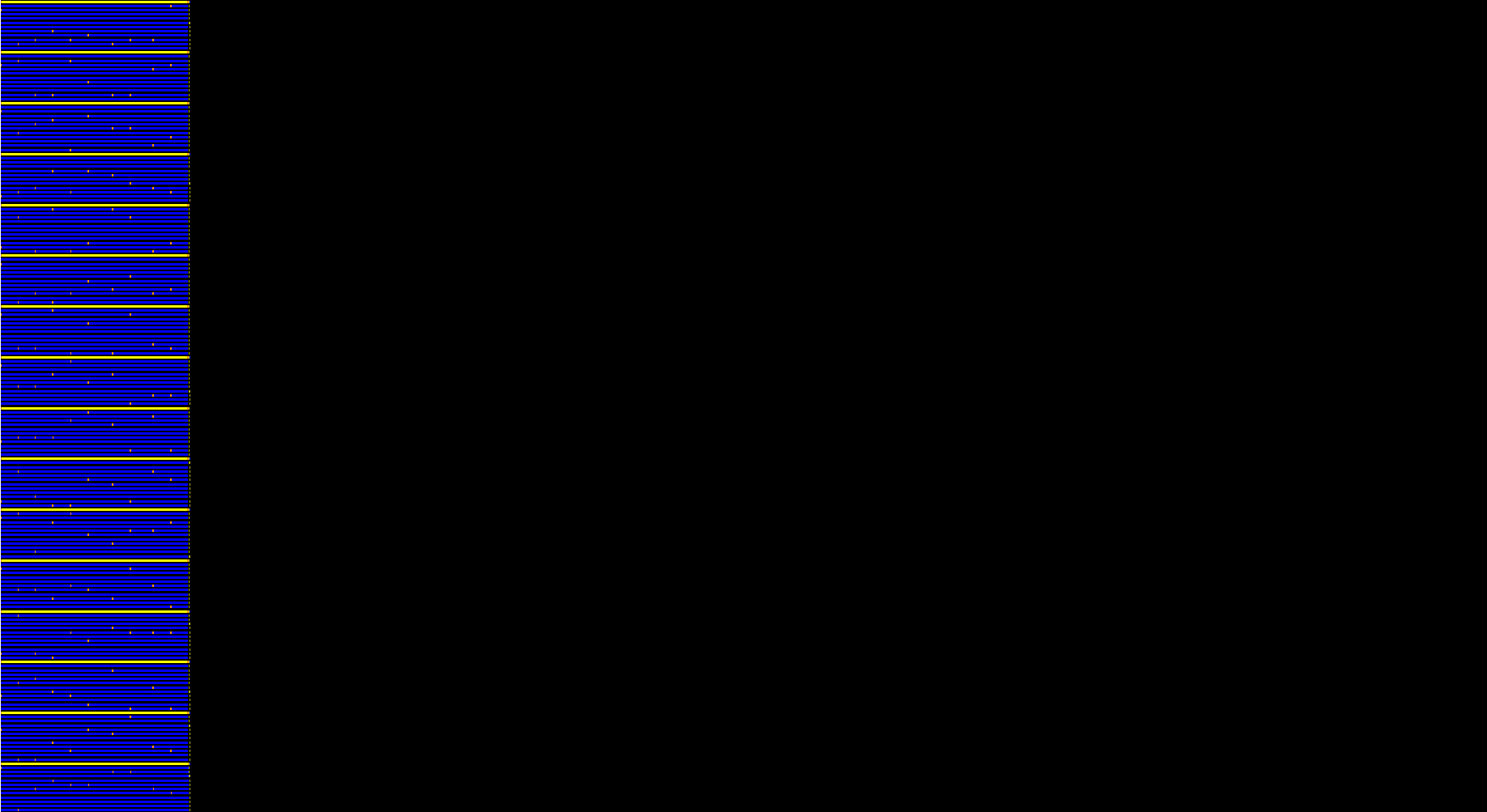}
\caption{Trace of a simulation of 10,000 neurons on 2 nodes (top) and 16 nodes (bottom). Horizontal axis represents the time consumed by the application, and the vertical axis represent the number of cores used.}
\label{p:fig5}
\end{figure*}

The first two traces (Figure~\ref{p:fig5}) correspond to the execution of 10,000 neurons on 2 and 16 nodes, respectively. Both traces have the same time-scale to see the differences in time. In the traces, we can see three different colors which correspond to \emph{OpenMP} scheduling (in yellow), \emph{OpenMP} execution (in blue) and \emph{MPI} communication (in orange). We see as many lines as cores used, 24 and 192 rows/cores for 2 and 16 nodes respectively (see Table~\ref{tab.platform1}). Using \emph{Extrae+Paraver} we are able to visualize easily the reduction in time achieved by increasing the number of nodes. In the 2-node trace (Figure~\ref{p:fig5}-top), it is difficult to see the orange color (\emph{MPI} communication), since the time consumed by \emph{MPI\_AllGather} is very low with respect to the time needed by the computation of the neurons just using 2 nodes.  However, when using 16 nodes (Figure~\ref{p:fig5}-bottom), it is easier to identify where the \emph{MPI} communication is performed. Increasing the number of nodes, we increase the complexity of the \emph{MPI} communication, the number of \emph{MPI} calls, and then the time consumed by these calls is longer. However, even when the use of \emph{MPI\_AllGather} poses an increment in time when using a higher number of nodes, the time consumed by these calls is less than 0.4\% of the total execution time. Furthermore, these calls are overlapped (using \emph{OpenMP} tasking) with the \emph{OpenMP} execution, so that this increment does not affect scalability. Only in extreme cases where we have a very low number of neurons per node (see Figure~\ref{p:fig2}, 100 neurons using 32 nodes) the time of the \emph{MPI} communication can affect scalability.        
Due to the \emph{OpenMP} scheduler, the MPI communication is done in one of the cores in each node, and the core responsible of the communication can change along the simulation.
In both traces the predominant color is the blue (\emph{OpenMP} computation). 

\begin{figure*}[t]
\centering
\includegraphics[width = 0.45\textwidth]{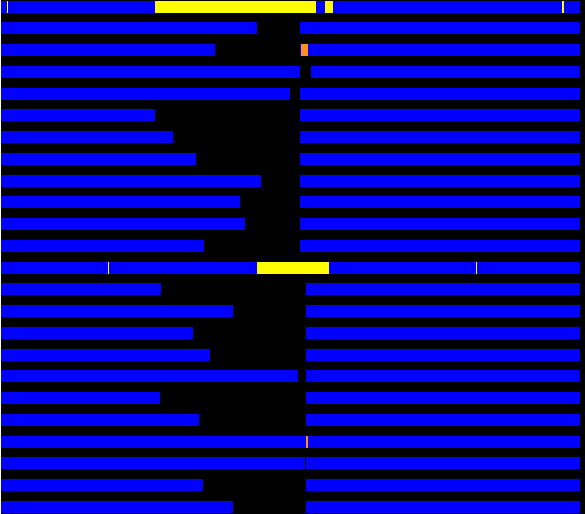}
\includegraphics[width = 0.45\textwidth]{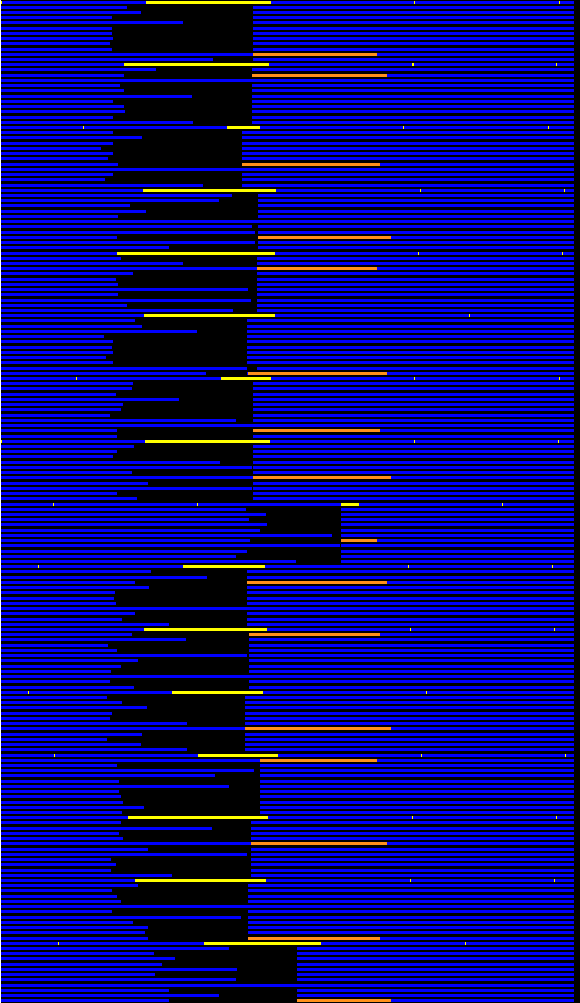}
\caption{Zoom in between the end of one global step and the beginning of the following global step for a simulation of 10,000 neurons on 2 nodes (left) and 10,000 neurons on 16 nodes (right).}
\label{p:fig6}
\end{figure*}


To analyze the \emph{MPI} communication and \emph{OpenMP} scheduling deeper, we zoom in between the end of one global step and the beginning of the following global step for both traces (Figure~\ref{p:fig6}). As in the previous traces, the time-scale is the same in these two traces. We can see more clearly the difference in time for \emph{MPI} communication. 
As expected, the \emph{MPI} calls are more time consuming by using 16 nodes than using 2 nodes. Anyway this is still low with respect to the computing time and, as commented before, the \emph{MPI} calls are overlapped with computation thanks to \emph{OpenMP} tasking. Hence, this increment does not affect scalability. Also, it is important to note that, unlike what we see in the previous traces (Figure~\ref{p:fig5}), the first cores of each node also compute some operations on the neurons, so they are not only busy computing \emph{OpenMP} instructions.

\section{Parallel MPI+OpenMP Tasking Simulator on \\ Multi-Morphology Simulations}
\label{parallel-multi}

Once the mono-morphology simulations have been deeply analyzed in the previous section, in this section we extend this analysis to multi-morphology simulations.
As most of the details about the parallelization implemented have been already presented, we start evaluating the impact of computing multi-morphology simulations on performance.
The first test consists of changing the size of the neurons by using a group of 20 neurons with different sizes initialized randomly (the maximum difference between the largest size and the smallest size is 700). We use this group to assign the size of 10,000 different neurons, so multiple neurons still have the same size. In order to evaluate the potential overhead that the multi-morphology simulations could suppose with respect to the mono-morphology configurations, we also execute a similar test case, but using just one size for all the neurons of the simulation. Both, the multi-morphology and the mono-morphology tests, present the same computational cost in terms of average size of the neurons (300).
We use 4 nodes of the same platform presented in the previous section.

\begin{figure*}[h]
\centering
\includegraphics[width = 0.7\textwidth]{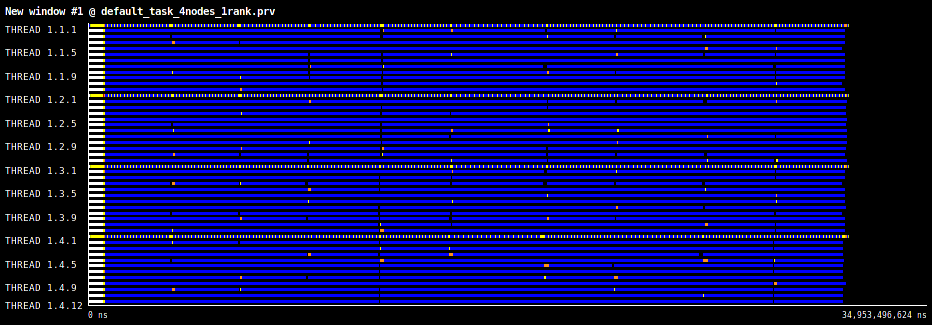}
\includegraphics[width = 0.7\textwidth]{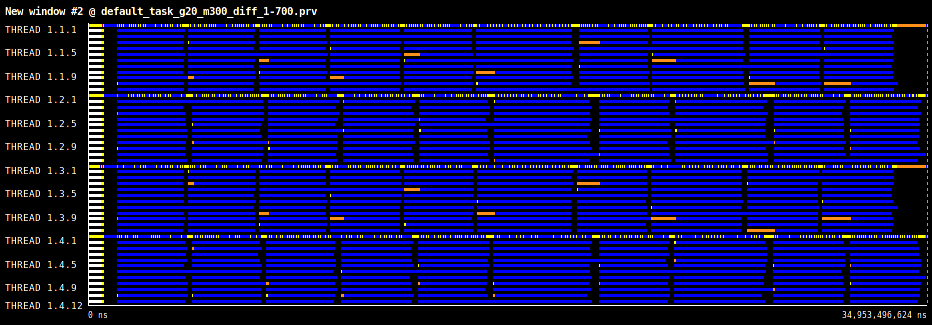}
\includegraphics[width = 0.7\textwidth]{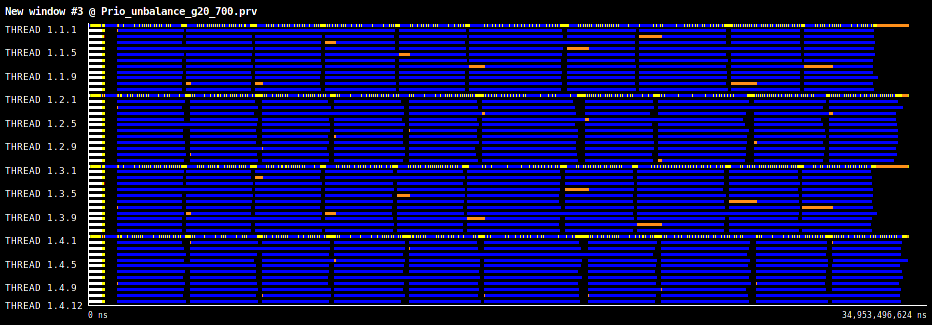}
\includegraphics[width = 0.7\textwidth]{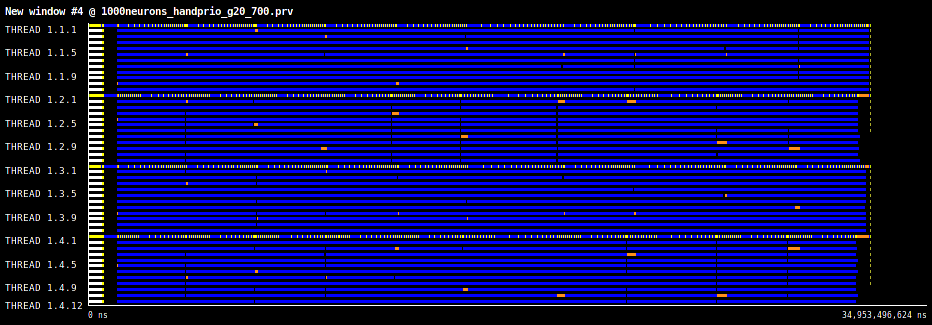}
\caption{Traces of a simulation of 10,000 neurons on 4 nodes for the mono-morphology test (first trace), the multi-morphology test (second trace), the implementation with \emph{OpenMP} priorities for the multi-morphology test (third trace), and the implementation with the initial sorting of the neurons (fourth trace).}
\label{pm:fig1}
\end{figure*}

Similarly to the previous section, we make use of the Extrae + Paraver tools to carry out the performance analysis. Figure~\ref{pm:fig1} illustrates the traces for the mono-morphology (first trace) and multi-morphology (second trace) tests. As shown, a few changes in terms of size of the neurons have important consequences on performance, the multi-morphology test being about 12\% slower in only 10 global iterations, even when both tests have the same computational cost in terms of average size of the neurons.   
We can note that the \emph{MPI} communication is longer in the multi-morphology test. 
Contrary to what one might think, this is not because of an increase in the amount of data  transferred in the MPI calls for the multi-morphology test. 
In fact, the data transfered in the \emph{MPI} communications is quite similar in both cases, mono-morphology and multi-morphology, and the number of \emph{MPI} calls are exactly the same. 
This unbalancing and the long \emph{MPI} calls presented in the trace corresponding to the multi-morphology test is mainly because of the unbalancing of the computational cost per node found in this kind of simulations. 

The first approach that we propose and analyze to minimize the unbalancing found between nodes consists of using \emph{OpenMP} priorities to execute the large neurons before the small neurons. This approach is not in need of important changes in the code. Basically, it consists of using the \emph{priority(x)} clause when instantiating the \emph{OpenMP} tasks, which compute the \emph{Hines} algorithm and the \emph{Spike} computation on the neurons. 
The priority assigned is the size of the neuron (see Algorithm~\ref{Pimplementation-priority}). 

\begin{algorithm}[h]
\caption{Parallel implementation based on MPI + OpenMP tasking + OpenMP priorities.}
\label{Pimplementation-priority}
\begin{small}
\begin{algorithmic} [1]
\State while(global\_step $\leq$ total\_steps)\{
\State \hspace{0.5cm}\#pragma omp parallel\{ 
\State \hspace{0.5cm}\#pragma omp master\{
\State \hspace{1cm}\#pragma omp task
\State \hspace{1cm}\emph{MPI\_AllGather(Spikes);}
\State \hspace{1cm}\#pragma omp task
\State \hspace{1cm}for( local\_step=0; local\_step $\leq$ D; local\_step++)\{
\State \hspace{1.5cm}for( i=0; i $\leq$ \#Neurons; i++)\{
\State \hspace{2cm}{\bf \#pragma omp task priority(sizeof(Neuron[i]))\{}
\State \hspace{2cm}\emph{Hines(Neuron[i]);}
\State \hspace{2cm}\emph{Spikes(Neuron[i]);}
\State \hspace{2cm}{\bf\} } //End omp task
\State \hspace{1.5cm}\} //End for \#Neurons
\State \hspace{1.5cm} \#pragma omp taskwait
\State \hspace{1cm} \} //End for total\_local\_steps 
\State \hspace{0.5cm}\} //End omp master
\State \hspace{0.5cm}\} //End omp parallel
\State \hspace{0.5cm} global\_step++;
\State \}
\end{algorithmic}
\end{small}
\end{algorithm}

\begin{figure*}[h]
\centering
\includegraphics[width = 0.65\textwidth]{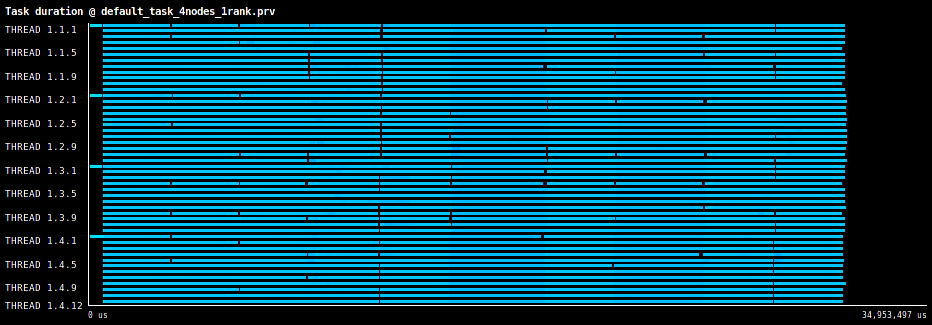}
\includegraphics[width = 0.65\textwidth]{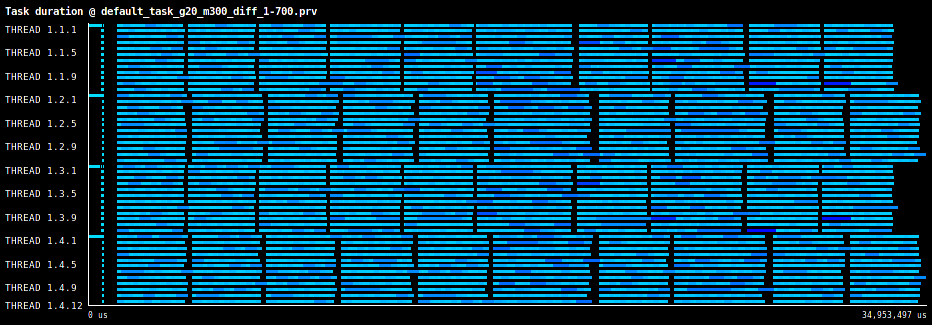}
\includegraphics[width = 0.65\textwidth]{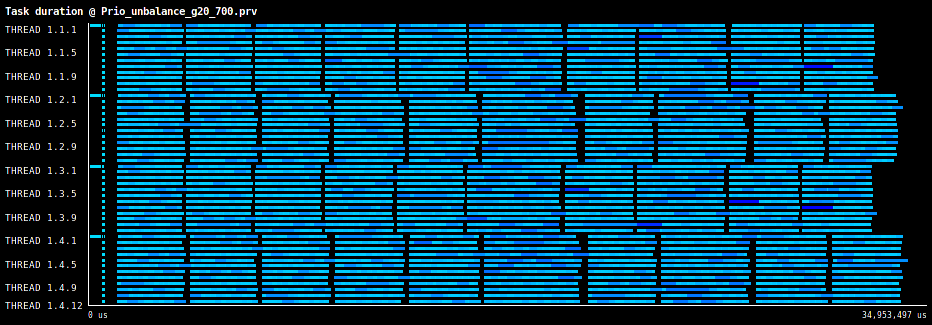}
\includegraphics[width = 0.65\textwidth]{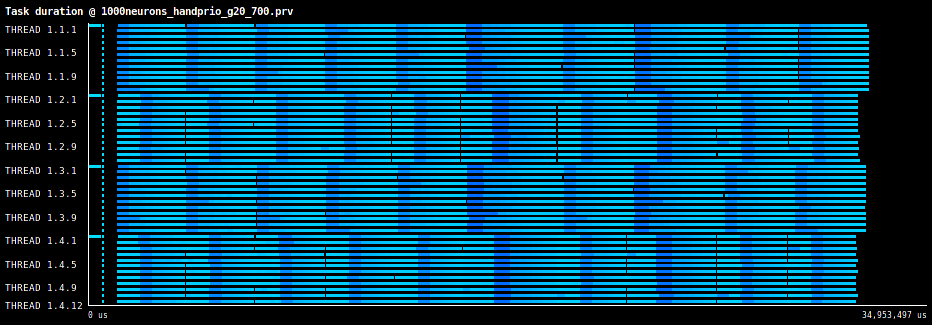}
\caption{Traces of a simulation of 10,000 neurons on 4 nodes with a different color diagram (the darker the color, the longer execution time) for the mono-morphology test (first trace), the multi-morphology test (second trace), the \emph{OpenMP} priorities for the multi-morphology test (third trace), and the implementation with the initial sorting of the neurons (fourth trace).}
\label{pm:fig2}
\end{figure*}

In order to evaluate the effectiveness of this approach deeper, Figure~\ref{pm:fig2} graphically illustrates the same traces presented before, but using a different color diagram, which help us to visualize and check if the big tasks (neurons) are effectively computed before the smaller ones. For that purpose, and for the sake of clarity, we selected that the darker the color, the longer the execution time.   

As shown in both figures (third trace of Figures~\ref{pm:fig1} and~\ref{pm:fig2}), the use of \emph{OpenMP} priorities does not achieve the expected results and it is not able to execute the large neurons (tasks) before the small neurons (tasks). Although this approach is easy to implement, the size of \emph{OpenMP} throttle~\cite{Thierry18}, which depends of the implementation, does not allow us to exploit the potential benefit of using priorities on our specific application, where we have a large number of tasks (as many task as neurons). The performance achieved is basically as fast as the version, which does not use priorities. 
Also, the data locality cannot be efficiently exploited for those groups of neurons where large neurons have been stored in memory very separated from each other. These results are in agreement with the results presented in~\cite{Valero-LaraPDP18}. As we will see in the rest of this section, the way that the neurons are stored in memory has very important consequences on performance.

Since the use of \emph{OpenMP} priorities does not present the effectiveness that we expected for our problem, we propose a different approach, which basically consists of sorting the neurons before the simulation according to their size. This can help not only to have a better balancing between nodes, but also a better data-locality in the execution on each of the nodes. 
As we can see (fourth trace of Figure~\ref{pm:fig1} and Figure~\ref{pm:fig2}), this simple change has important benefits. The time consumed is quite similar to the time of the mono-morphology test, being only 2,8\% slower. This is an important reduction with respect to the original multi-morphology test. As shown, the size of the \emph{MPI} calls were reduced due to a better scheduling. However, the large neurons are not executed before the small counterparts in all the nodes (see fourth trace of Figure~\ref{pm:fig2}); this is because of the sorting of the neurons and the assignation of these to the computational nodes.

To analyze the performance and the efficiency of our approach on more realistic scenarios, we perform one last test assigning the size of the neuron according to an F-Fisher distribution. Multiple studies have proven that the size of the neurons follows this kind of statistical distributions~\cite{Paolo13}. We follow the same sorting done in the previous test. For the sake of simplicity, let us call this sorting as \emph{global sorting-storing}.    

\begin{figure*}[h]
\centering
\includegraphics[width = 0.95\textwidth]{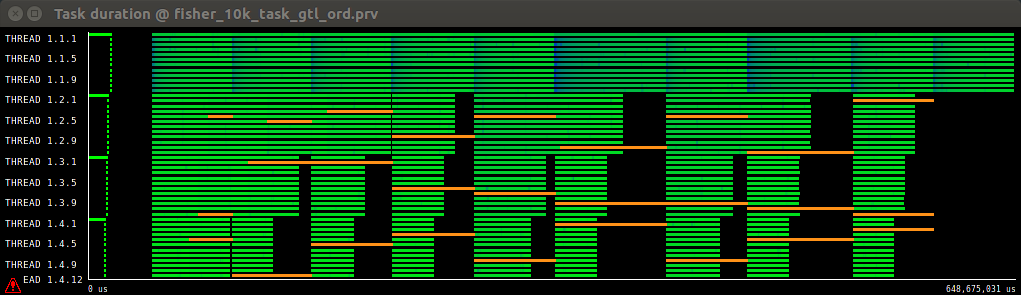}
\includegraphics[width = 0.95\textwidth]{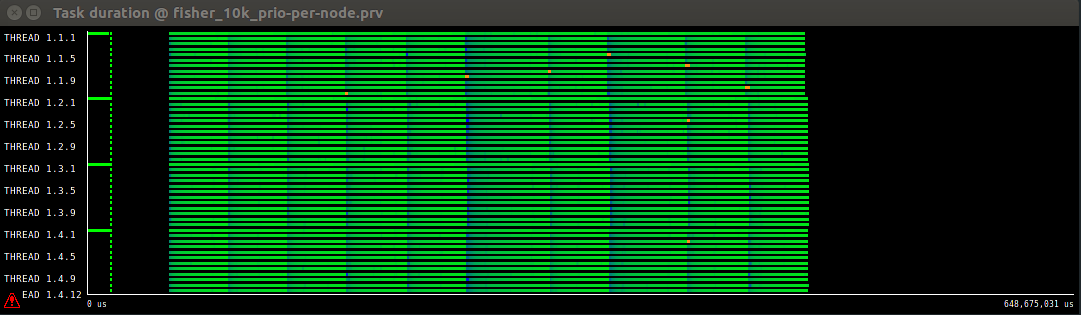}
\caption{Traces of a simulation of 10,000 neurons on 4 nodes for the multi-morphology test using as size of the neurons a F-Fisher distribution using \emph{global sorting-storing} (top) and \emph{local sorting-storing} (bottom).}
\label{pm:fig5}
\end{figure*}

In the last tests analyzed, we realized that the sorting and storing of the neurons on the different computational nodes can influence substantially the execution time of the simulation, as well as the efficient use of our computational resources. Therefore, it is of vital importance the way in which we distribute the group of neurons between the nodes.
We want to confirm this by using a more realistic neural network. We have used the same number of nodes and number of neurons than in the previous tests. 
However, we have used an F-Fisher distribution to assign the size of the neurons.
Two different approaches are tested. The first consists of sorting all the neurons globally, so the largest neurons are stored in the first of the four nodes, and the smallest neurons in the last node.
As expected, this approach causes an important underutilization of the computational resources, where an important part of these resources is not used along the simulation. Even in the last global step, only the first node is computing, the rest of nodes being completely idle (see Figure~\ref{pm:fig5}-top).
This is mainly due to two consequences: first, the unbalancing between nodes; second, the communication among nodes which, as we introduced in Section~\ref{simulator}, is an important synchronization point between global steps, this being the cause behind the large communication tasks (orange lines) in Figure~\ref{pm:fig5}-top.

To solve this problem, we propose a different approach. As commented before, the size of the neurons follows a particular statistical distribution depending on the brain area of interest~\cite{Paolo13}.
We generate as many groups of neurons as number of nodes, and then we sort these groups independently in each of the nodes (\emph{local sorting-storing}). The neurons generated in each of the nodes are sorted to achieve a good balancing not only between nodes but also between cores of the same node.
This is mainly because of the use \emph{OpenMP} tasking. Using tasks and sorting the group of neurons in each node(\emph{local storing-storing}), we are able to balance the workload between nodes and cores, minimizing considerably the time of the simulation (about 30\% of the time), by using efficiently all the resources available (see Figure~\ref{pm:fig5}-bottom).



It is also important to note that this preprocessing (sorting and storing) is computationally expensive, but this has to be computed just once at the very beginning of the simulation, so the time consumed by this preprocessing is negligible when compared to the hours of computation that these simulations need.

\section{Related Work}
\label{related}

It is possible to find many initiatives for the simulation of the behavior of the human brain by computers across the world, for instance in the USA~\cite{BRAIN}, Europe~\cite{HBP, Blue, NEST} and Japan~\cite{OKANO2016,BRAINMIDS}. Each of these initiatives is focused on the development of a set of tools for such target. 
In particular, we have focused on one of the tools developed in the European initiative and on the scalability study using \emph{MPI+OpenMP} tasking. 
 
Most of the state-of-the-art references are focused on accelerating one of the most computationally expensive steps, that is the voltage capacitance on neurons' morphologies.
The standard algorithm used to compute the voltage on neurons' morphology is the Hines algorithm~\cite{Hines84}.
This algorithm is based on the Thomas algorithm~\cite{Conte80}, which solves tridiagonal systems. 
Although the use of GPUs to compute the Thomas algorithm has been deeply studied~\cite{Valero-LaraPPAM17, Valero-LaraCPE18, Valero-Lara14, Valero-Lara12, Davidson11, Zhang10, cuSPARSE}, the differences
among these two algorithms, Hines and Thomas, makes us impossible to use the latter, since it cannot deal with the sparsity of the Hines matrix. 
Recently, a new methodology was proposed to deal with a very high number of neurons on NVIDIA GPUs, achieving good scalability~\cite{Valero-LaraICCS17,Valero-LaraOGTS18}.

Unlike previous works, we focused on the parallelization and scalability on the \emph{whole} application based on \emph{MPI+OpenMP} tasking. In fact, the use of \emph{MPI+OpenMP} tasking has been also used in other applications~\cite{Marjanovic2010}, such as the HPLinpack~\cite{HPL} tool used for the performance analysis of the TOP-500 list~\cite{TOP500}, achieving good results.

\section{Conclusions and Future Work}
\label{conclusions}

In the present work the authors proposed and evaluated an efficient (in terms of programmability) and optimized (in terms of performance and scalability) implementation 
for one of the most important challenges into the scientific community today, that is the simulation of the human brain.
Given the results obtained and presented in this paper, the authors have proven the efficiency of using \emph{MPI+OpenMP} tasking to achieve good scaling.

Although \emph{MPI\_AllGather} and \emph{OpenMP} tasking can cause an important infra-utilization of the computational resources, if a good scheduling is not used or there is a bad distribution of
the workload among nodes, it has been proven that this approach is not only an affordable and reduced-cost implementation in terms of programmability, but also it is able to exploit 
efficiently the strategy implemented in the simulator, achieving even an ideal scaling. However, it is of vital importance to balance the workload between nodes; otherwise, the
effectiveness of this model is not well exploited, performing an important infra-utilization of the computational resources, as in the case of multi-morphology simulations.

Multi-morphology simulations present an important challenge in terms of scheduling to achieve a balanced workload distribution between nodes. 
In fact, an important underutilization of the resources is presented, even in those simulations where the neurons, in terms of size, are not very different between them.
We have proven that the way in which the neurons are distributed on the different nodes has important consequences to achieve a good performance for this kind of simulations.
Unfortunately, the use of \emph{OpenMP} priorities does not obtain the expected results for a better balancing between nodes. 
However, storing the large neurons before the smaller, and following this same order when accessing to memory can alleviate the important underutilization of the computational resources.  
To analyze this approach on more realistic scenarios, we have followed an F-Fisher distribution to assign the size of the neurons. 
In this case, all the neurons are completely different
between them, which causes a much more important underutilization of computational resources and unbalancing between nodes.
To solve this problem, we propose: i) to distribute the creation of the neurons in each of the nodes and ii) storing the neurons regarding their size (the large neurons before the smaller) in each of the nodes. This reduces considerably the execution time of the simulation by improving the usability of the computational resources and the balancing between nodes. In fact, the behavior achieved when this approach is used is quite similar to the results reported for mono-morphology simulations.

\section*{Acknowledgement}
We would like to appreciate the valuable feedback and help provided by the main developers of the Arbor simulator: Benjamin Cumming (ETH Z\"urich) and Alexander Peyser (J\"ulich Supercomputing Center).
This project has also received funding from the European Union's Horizon 2020 research and innovation programme under grant agreement No 720270 (HBP SGA1), from the Spanish Ministry of Economy and Competitiveness under the project Computaci\'on de Altas Prestaciones VII (TIN2015-65316-P) and the Departament d'Innovaci\'o, Universitats i Empresa de la Generalitat de Catalunya, under project MPEXPAR: Models de Programaci\'o i Entorns d'Execuci\'o Paral$\cdot$lels (2014-SGR-1051). 
Finally, this project also received funding from the Spanish Ministry of Economy and Competitiveness under the Juan de la Cierva Grant Agreement No IJCI-2017-33511, and from the European Union's Horizon 2020 research and innovation program under the Marie Sklodowska Curie grant agreement No. 749516. 

\bibliographystyle{elsarticle-num}
\bibliography{biblio}

\end{document}